\documentclass{ifacconf}
\usepackage{graphicx}      
\usepackage{natbib}
\usepackage{graphics} 
\usepackage{amsmath,amsfonts} 
\usepackage{amssymb}  
\usepackage{color,xcolor}
\usepackage{comment}
\newtheorem{lemma}{Lemma}
\begin{document}
\begin{frontmatter}
\title{Solvability of the Output Corridor Control Problem  by Pulse-Modulated Feedback
}

\author[Second]{Alexander Medvedev}
\author[First]{Anton V. Proskurnikov}

\address[Second]{Department of Information Technology,
       Uppsala University, SE-752 37 Uppsala, Sweden (e-mail: alexander.medvedev@it.uu.se)}
\address[First]{Department of Electronics and Telecommunications, Politecnico di Torino, Turin, Italy, 10129  (e-mail: anton.p.1982@ieee.org).}



\begin{abstract}
The problem of maintaining the output of a  positive time-invariant single-input single-output system within a predefined corridor of values is treated. For third-order plants possessing a certain structure, it is proven that the problem is always solvable under stationary conditions by means of pulse-modulated feedback. The obtained result is utilized to assess the feasibility of patient-specific pharmacokinetic-pharmacodynamic models with respect to patient safety. A population of  Wiener models capturing the dynamics  of a neuromuscular blockade agent is studied to investigate whether or not they can be driven into the desired output corridor by clinically acceptable sequential drug doses (boluses). It is demonstrated that low values of a parameter in the nonlinear pharmacodynamic part lie behind the detected model infeasibility.
\end{abstract}
\end{frontmatter}
\section{Introduction}
 Classical digital control addresses the problem of capturing essential continuous  plant dynamics in a discrete model through the choice of the sampling time, often by applying engineering rules of thumb~\citep{AAW13}. Then, the controller design is carried out utilizing the discretized model. Combined with model-based control algorithms, the necessity of high sampling rates resulted in significant computational burden, posing an implementation challenge to the control hardware of that time. Taking into account the inter-sample behavior in controller design emerged as an effective way of reducing the sampling rate to a minimum, an approach supported by sampled-data control theory~\citep{T92}.

 The advances in the performance of control computers made the computation cheap and exceedingly high sampling rates readily available, thus rendering the sampled-data control tools inopportune. Meanwhile, the transition from centralized to networked control highlighted the issues related to irregular sampling that stem from network imperfections, e.g., delays, packet loss, and jitter. These problems can be properly addressed by allowing irregular sampling in the framework of event-based control~\citep{AA08}. Depending on the application, event-based control can reduce sampling rates by orders of magnitude, achieving similar performance to that of periodic sampling, while saving bandwidth and computation.

 Asynchronous sampling of event-based control is still aimed at approximating the performance of regular discrete controllers, although with a lower communication demand. Some control applications do not require tight feedback and allow for episodic adjustments of the control signal. A typical example of such an application is dosing, that is, the process of administering a measured amount of a substance, which is often performed in a discrete manner.

A commonplace example of discrete dosing is following a doctor's orders on a medication regimen. Drug dosing often constitutes a repetitive and periodic activity where the doses and inter-dose intervals are adjusted to achieve the desired result and minimize the medication side effects. Interpreting administration of individual doses as impulsive action and allowing for measurement of the therapeutical effect directly leads to the concept of pulse-modulated feedback~\citep{GC98}, where doses are manipulated through amplitude modulation and inter-dose intervals through frequency modulation. Yet, impulsive (or pulse-modulated) control is seldom used in the practice of drug dosing. Instead, standard process control techniques exploiting PID-controllers and model-predictive controllers (MPC) dominate the area.

MPC is also possible in impulsive drug dosing setups. An application of impulsive MPC to control the intravenous bolus administration of lithium is reported in~\cite{SPS15}. A promising application of impulsive MPC is insulin dosing in simulated diabetes patients~\citep{RGS20,EBC23}. Impulsive control, even when applied to linear systems, introduces strongly nonlinear dynamics and, consequently, inflicts significant challenges in stability and robustness analysis; online optimization further complicates the situation.

A simple pulse-modulated controller that mimics pulsatile endocrine regulation and is suitable for discrete dosing applications has been suggested in~\cite{MPZ24}. Further study of the closed-loop dynamics arising when the controller is employed to the dosing of a neuromuscular blockade agent revealed complex nonlinear phenomena such as periodic solutions of high multiplicity, bistability, and deterministic chaos~\citep{MPZ25}. Bifurcation analysis is proposed as a way of avoiding undesired behaviors. Of course, it requires accurate patient-specific modeling of the pharmacokinetics and pharmacodynamics (PKPD), which is difficult to obtain from clinical data.

The contribution of this paper is twofold: First, it is proven that, for a certain class of third-order models, the problem of keeping the stationary plant output within a predefined corridor of values is always solvable by means of pulse-modulated feedback. Second, based on the obtained controller design insights, a method for assessing the feasibility of patient-specific PKPD models with respect to patient safety is proposed and illustrated on a previously published cohort~\citep{SWM12} of patient-specific models estimated from clinical data.

The rest of the paper is organized as follows. Section~\ref{sec:problem} defines the mathematical model at hand and formulates the output corridor impulsive control problem. Section~\ref{sec:background} provides background information on necessary mathematical tools and previously proven results. Section~\ref{sec:results} presents the main results of the paper; a sketch of the proof is given in the final Section~\ref{sec:proof}. Section~\ref{sec:application} describes the application to feasibility assessment of patient-specific PKPD models.

\section{Problem definition}\label{sec:problem}

Consider a continuous-time linear time-invariant system
\begin{equation}                            \label{eq:state-space}
\dot{x}(t) =Ax(t)+Bu(t), \quad y(t)=Cx(t),
\end{equation}
driven by a train of impulses,
\begin{equation}                            \label{eq:pulses}
u(t)=\sum\nolimits_{n=0}^{\infty}\lambda_n\delta(t-t_n),\quad 0=t_0<t_1<t_2<\ldots,
\end{equation}
where $\delta(\cdot)$ is Dirac delta-function.
The system matrices have the following structure:
\begin{equation}\label{eq:matrices}
A=\begin{bmatrix} -a_1 &0 &0 \\ g_1 & -a_2 &0 \\ 0 &g_2 &-a_3 \end{bmatrix}, B=\begin{bmatrix} 1 \\ 0 \\ 0\end{bmatrix}, C =\begin{bmatrix} 0 &0 &1 \end{bmatrix},
\end{equation}
with distinct $a_1,a_2,a_3>0$ and $g_1,g_2>0$. Since the matrix $A$ is Metzler, system~\eqref{eq:state-space} is positive, that is, $x(t) \geq 0$ for all $t \in [0,\infty)$ whenever $x(0)\geq 0$ and $\lambda_n\geq 0$.

Mathematically,  system~\eqref{eq:state-space} under  control law~\eqref{eq:pulses} constitutes a hybrid system, which evolves according to a continuous flow between the consecutive impulses,
\begin{equation} \label{eq:1a}
\dot{x}(t) = Ax(t), \qquad y(t) = Cx(t), \quad \forall t \in (t_n, t_{n+1}),
\end{equation}
and undergoes instantaneous jumps in the state vector when an impulse occurs\footnote{The superscripts $-$ and $+$ in~\eqref{eq:1b} denote the left-sided and right-sided limits, respectively.}:
\begin{equation} \label{eq:1b}
x(t_{n}^+) = x(t_{n}^-) + \lambda_n B, \qquad n = 0,1,\ldots
\end{equation}

Feedback is enforced by modulating jump timing and magnitude with the continuous output,
\begin{equation}\label{eq:modulation}
    T_n =\Phi( y(t_n)), \quad \lambda_n=F( y(t_n)), \quad t_{n+1}= t_n+T_n.
\end{equation}
The recursion above makes the controller dynamical of first order.
The design degrees of freedom of pulse-modulated controller~\eqref{eq:modulation} are the frequency modulation function $\Phi(\cdot)$ and the amplitude modulation function $F(\cdot)$ that can be selected to achieve a variety of control objectives.
To guarantee that the solutions are positive and bounded, it suffices to assume that these functions are bounded from above and below:
\begin{equation}                             \label{eq:2a}
0<\Phi_1\le \Phi(\cdot)\le\Phi_2, \quad 0<F_1\le F(\cdot)\le F_2,
\end{equation}
where $\Phi_1$, $\Phi_2$, $F_1$, $F_2$ are constants.

To summarize the operation of the impulsive feedback controller at each firing time $t_n$, it reads out the system output $y(t_n)$ and calculates the next firing instant $t_{n+1}$ using the frequency modulation function, as well as the impulse weight $\lambda_n$ using the amplitude modulation function.

 The following control problem is treated  in this paper.

\paragraph*{\bf Output corridor impulsive control problem:}
Given  plant~\eqref{eq:state-space},
design the modulation functions $\Phi(\cdot)$
and $F(\cdot)$ of  impulsive feedback law~\eqref{eq:1b},~\eqref{eq:modulation} that  maintain the steady-state output $y(t)$ in the pre-defined corridor
\begin{equation}\label{eq:corridor}
    y(t)\in \lbrack y_{\min}^*,y_{\max}^*\rbrack, \quad y_{\min}^*>0.
\end{equation}

While the solvability of this general problem is nontrivial due to the diversity of complex nonlinear phenomena in the closed-loop dynamics, we study its particularization: when does a special periodic solution termed a 1-cycle (see the definition in the next section) and satisfying~\eqref{eq:corridor} exist? This problem is important for two reasons: first, 1-cycles are the simplest periodic solutions of the hybrid system (one impulse over the least period); second, a design procedure leading to a stable 1-cycle with predefined parameters~\citep{MPZh23,MPZh23a,MPZ24} readily exists.

\section{Background}\label{sec:background}

\paragraph*{\bf Reduction to discrete-time dynamics:}

It is easily seen that the trajectory of the hybrid closed-loop dynamics~\eqref{eq:1a}-\eqref{eq:modulation} is completely determined by the sequence $X_n = x(t_n^-)$, which evolves according to
\begin{equation} \label{eq:1c}
X_{n+1} = \e^{(t_{n+1}-t_n)A}(X_n + \lambda_n B), \qquad n = 0,1,\ldots
\end{equation}
where $t_{n+1}-t_n$ and $\lambda_n$ are determined by~\eqref{eq:modulation}. Indeed, given $X_n$,  the trajectory of~\eqref{eq:1a} on the interval $(t_n,t_{n+1})$ is \begin{equation} \label{eq:1d}
x(t)=\e^{(t-t_n)A}(X_n+\lambda_n B),\quad t\in(t_n,t_{n+1}).
\end{equation}
Therefore, the sequence $X_n, n = 0,1,\ldots$ completely defines the solution of the hybrid system.
\paragraph*{\bf 1-cycles and their characteristics:} A periodic solution of \eqref{eq:1a}-\eqref{eq:modulation} is called a 1-cycle if there is only one impulse firing of the pulse-modulated feedback within the least period, i.e., $\lambda_n\equiv\lambda$, $T_n\equiv T$ for all $n=0,1,\ldots$.
 Then, a 1-cycle corresponds to a fixed point of the discrete map
 \begin{equation}\label{eq:1-cycle}
    X=Q(X), \quad Q(\xi)\triangleq\mathrm{e}^{A\Phi(C\xi)}\left( \xi+ F(C\xi)B \right).
\end{equation}
Assuming that $F$ is nonincreasing and $\Phi$ is nondecreasing (the hybrid system in this case turns into the impulsive Goodwin's oscillator), the 1-cycle exists and is unique~\citep{Aut09,PRM24}.

However, solving the  output corridor impulsive control problem requires to design a 1-cycle with given parameters through the choice of appropriate modulation functions. It appears~\citep{MPZh23} that the fixed point is uniquely determined by the 1-cycle parameters $(\lambda, T)$
\begin{equation}\label{eq:X-explicit}
X = \lambda(\e^{-AT} - I)^{-1}B.
\end{equation}
Indeed, for a periodic solution to~\eqref{eq:1a}-\eqref{eq:modulation} with $X_n\equiv X$,  $\lambda_n\equiv \lambda=F(CX)$ and $T_n\equiv T=\Phi(CX)$, equation~\eqref{eq:X-explicit}, obviously, is equivalent to $X=Q(X)$.

Notice that expression~\eqref{eq:X-explicit} can be further simplified as the matrix exponential of $A$
can be explicitly computed via the Opitz formula~\citep{DeBoor2005}.

Defining the first divided difference of a function $\psi$ as
\[
\psi[\xi_0,\xi_1]\triangleq \frac{\psi(\xi_1)-\psi(\xi_0)}{\xi_1-\xi_0},
\]
where $\xi_0\ne \xi_1$, the second divided difference is a function of three variables and is defined by
\[
\psi[\xi_0,\xi_1,\xi_2]\triangleq\frac{\psi[\xi_1,\xi_2]-\psi[\xi_0,\xi_1]}{\xi_2-\xi_0},
\]
where $\xi_0,\xi_1,\xi_2$ are pairwise distinct.

\begin{prop}{\cite[Proposition~2]{MPZh23}}\label{th:fp}
The positive fixed point $X=[X_{1},X_{2},X_{3}]^\top$ of the map $Q$ in~\eqref{eq:1-cycle} is uniquely determined by the parameters of the $1$-cycle, $T > 0$ (the period) and $\lambda > 0$ (the impulse weight).
In terms of individual elements, it reads
\begin{align}\label{eq:x0_elements}
     X_{1}&=\lambda\mu(-a_1T), \\
     X_{2}&=\lambda g_1T\mu[-a_1T,-a_2T],\nonumber\\
     X_{3}&=\lambda g_1g_2T^2\mu[-a_1T,-a_2T,-a_3T], \nonumber
 \end{align}
 where $\mu(x)\triangleq\frac{1}{\e^{-x}-1}$.
\end{prop}

\paragraph*{\bf The Output Corridor Problem for 1-Cycles:} We now focus on a special case of the general output corridor control problem: find the parameters of the  1-cycle $(\lambda,T)$ such that the output $y(t)=Cx(t)$, corresponding to the 1-cycle state $x(t)$ from~\eqref{eq:1c},~\eqref{eq:1d}, satisfies the corridor condition in~\eqref{eq:corridor}. Naturally, to guarantee (local) attractivity of the 1-cycle, it has to be orbitally stable~\citep{S09}.

To handle this problem, we need the following extremal properties of the output $y(t)$ in a 1-cycle. Notice that, for each $0\leq\tau\leq T$,
the output is determined by the fixed point~\eqref{eq:X-explicit} and given by
\begin{equation}\label{eq:1e}
y(\tau)=\lambda z(\tau),\quad z(\tau,T)\triangleq C\e^{A\tau}{(I-\e^{AT})}^{-1}B.
\end{equation}
Then the output is a linear function of the impulse magnitude $\lambda$ and a nonlinear function of the period $T$.
Using this representation, the following can be proved.

\begin{prop}\label{prop:corridor}
For every period $T>0$, the equation
\begin{equation}\label{eq:roots_dy}
  \frac{\partial z}{\partial\tau}(\tau,T)=C\e^{A\tau}{(I-\e^{AT})}^{-1}AB=0
\end{equation}
has exactly two roots $\tau_1$ and $\tau_2$, satisfying $0<\tau_1<\tau_2<T$. These roots correspond, respectively, to the minimum and the maximum of the system output $y(t)$, defined by~\eqref{eq:1e}, on $[0,T]$. The output $y(t)$ decreases on the intervals $(0,\tau_1)$ and $(\tau_2,T)$ and increases on $(\tau_1,\tau_2)$. The maximum and minimum of the output on $[0,T]$ are, respectively,
\begin{align}
    y_{\max}&=\lambda z(\tau_2,T)>X_3,\label{eq:y_max}\\
    0<y_{\min}&=\lambda z(\tau_1,T)<X_3. \label{eq:y_min}
\end{align}
\end{prop}

Due to the linearity of~\eqref{eq:1b}, the time instants $\tau_1,\tau_2$ when the output $y(t)$ achieves the extreme values are independent of $\lambda$, which, according to~\eqref{eq:X-explicit}, scales the fixed point of the 1-cycle. Remarkably, the minimum of the output is achieved strictly between the impulses, which is counterintuitive as  jumps~\eqref{eq:1b} increase the state vector. Nevertheless, the last coordinate $y=x_3=Cx$ is not immediately affected by these jumps, remaining decreasing for a positive time $\tau_1$ after the impulse.

Proposition~\ref{prop:corridor} has an important consequence for  the  manual administration of drugs whose PKPD models obey \eqref{eq:state-space}, \eqref{eq:matrices}. When it is observed that the measured output (i.e. drug concentration in the effect compartment) has crossed the lower bound (e.g. the minimum effective concentration), it will continue to decrease for a while even though a new drug bolus is administered at the very same moment when the observation was made.
Therefore, to keep the output within the desired corridor, either manual administration should be predictive or properly designed closed-loop drug administration should be employed.

\section{Main Result}\label{sec:results}

The main result of the paper is constituted by an affirmative answer to the Output Corridor Problem for 1-Cycles formulated in Section~\ref{sec:background}: for any prescribed positive output corridor, there exists a 1-cycle whose output satisfies the corridor constraint. Furthermore, it is shown that, for each range $[y_{\min}^*$, $y_{\max}^*]$,  there exist parameters $(\lambda^*,T^*)$ such that the corridor constraint is tight for the 1-cycle corresponding to the fixed point $X^*$, that is, $y_{\min}=y_{\min}^*$ and $y_{\max}=y_{\max}^*$.

With $\tau_1,\tau_2$ defined in Proposition~\ref{prop:corridor} and for the function $z$ introduced in~\eqref{eq:1e}, denote
\begin{align}
    z_{\max}(T)&\triangleq z(\tau_2,T)=C\e^{A\tau_{2}}{(I-\e^{AT})}^{-1}B, \label{eq:z_max}\\
    z_{\min}(T)&\triangleq z(\tau_1,T)=C\e^{A\tau_{1}}{(I-\e^{AT})}^{-1}B. \label{eq:z_min}
\end{align}
Here $\tau_1,\tau_2$ are the same as in Proposition~\ref{prop:corridor}.

\begin{figure}[htb]
    \centering
    \includegraphics[width=1\linewidth]{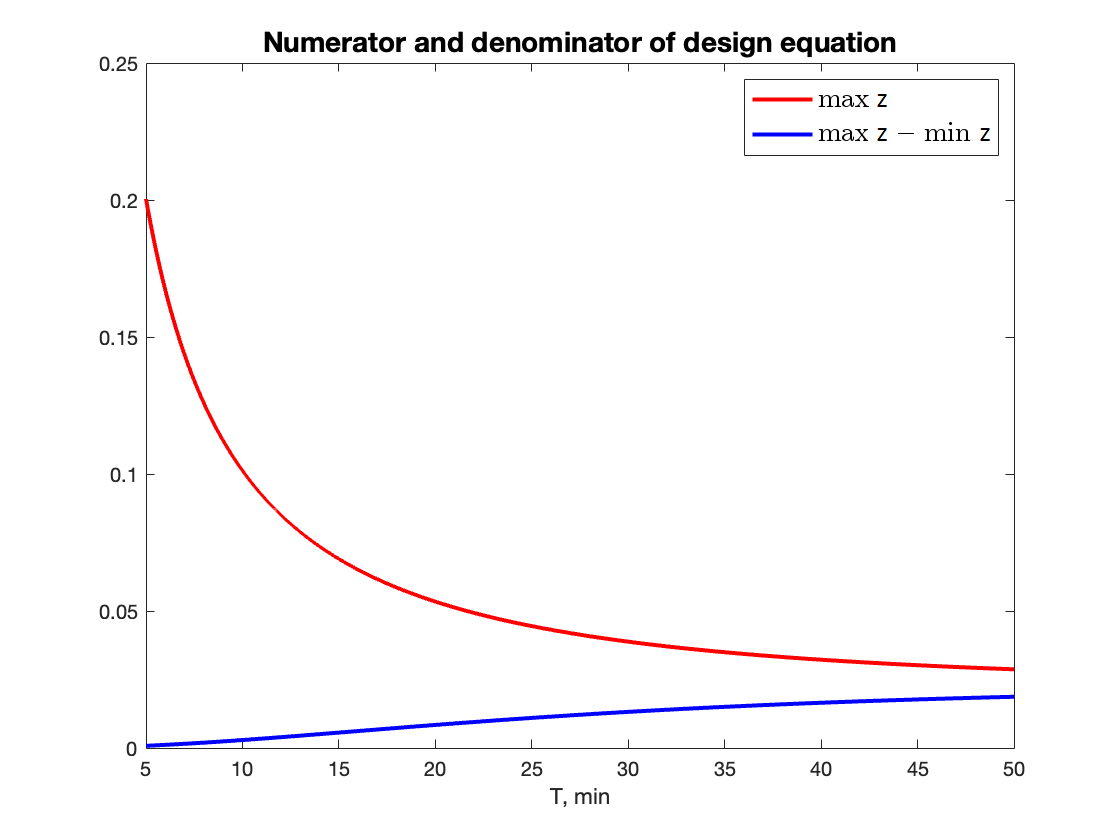}
    \caption{Numerator and denominator of the function $\Psi(T)$ in design equation \eqref{eq:design_T}. It can be seen that
    $\lim_{T\to\infty}\Psi(T)=1$, $\lim_{T\to 0}\Psi(T)=\infty$.}
    \label{fig:num_den_design}
\end{figure}

\begin{thm}\label{th:solvability}
Given output corridor~\eqref{eq:corridor} with $0<y_{\min}^*<y_{\max}^*$, the equation
\begin{equation}\label{eq:design_T}
\Psi(T) \triangleq \frac{z_{\max}(T)}{z_{\max}(T)\!-\!z_{\min}(T)} =\frac{y_{\max}^*}{ y_{\max}^*\! - \!  y_{\min}^*},
\end{equation}
always has a 
solution $T^*>0$. Denoting
\begin{equation}\label{eq:lambda}
    \lambda^*=\frac{y_{\max}^*}{z_{\max}(T^*)},
\end{equation}
the output $y(t)$ of \eqref{eq:1a}, \eqref{eq:1b} in the 1-cycle with parameters $(\lambda^*,T^*)$ satisfies~\eqref{eq:corridor}, and both inequalities are tight:
\[
y_{\min}=y_{\min}^*,\quad y_{\max}=y_{\max}^*.
\]
\end{thm}
\vskip0.1cm
The proof of Theorem~\ref{th:solvability} is given in Section~\ref{sec:proof}.
The key idea of the proof is illustrated by Fig.~\ref{fig:num_den_design}. The numerator of $\Psi(T)$ (that is, $z_{\max}(T)$) grows unbounded as $T\to 0$, i.e.,
$\lim_{T\to 0}z_{\max}(T)=+\infty$, while the denominator $z_{\max}-z_{\min}$ remains bounded as $T\to 0$.
Also, $z_{\max}(T)$ has a constant limit as $T\to\infty$, while $z_{\min}(T)$ vanishes. Hence, $\Psi(T)\to 1$ as $T\to\infty$, $\Psi(T)\to\infty$ as $T\to 0$, and $\Psi$ is continuous. 

\begin{rem}
Extensive simulations show that the function $\Psi$ is decreasing, as the numerator $z_{\max}$ and the denominator $z_{\max}-z_{\min}$ are, respectively, decreasing and increasing (Fig.~\ref{fig:num_den_design}). The complete proof of this fact remains an open problem, left for future research.
\end{rem}

\section{Model feasibility analysis}\label{sec:application}

\subsection{Model cohort}
Neuromuscular blockade (NMB) is a medical procedure that induces temporary skeletal muscle paralysis by interrupting nerve impulses at the neuromuscular junction. It is widely used in operating rooms and intensive care units to facilitate safe airway management, optimize surgical conditions, and manage critically ill patients on ventilators.

A minimal patient-specific PKPD Wiener model of the NMB agent \emph{atracurium} is introduced in~\cite{SWM12} and is intended for individualized control of drug delivery under general anesthesia.
Given typically limited excitation in drug dosing applications, minimizing the number of estimated parameters is necessary to achieve reasonable modeling accuracy.

The PK part of the model is given by the transfer function from the input $u(t)$ to the serum drug concentration $\bar y(t)$
\begin{equation}\label{eq:lin_NMB}
W(s)=\frac{\bar Y(s)}{U(s)}=\frac{v_1 v_2 v_3 \alpha^3}{(s+v_1\alpha)(s+v_2\alpha)(s+v_3\alpha)}.
\end{equation}
Here, $\bar Y (s)={\cal L}\{\bar y(t)\}$, $U (s)={\cal L}\{ u(t)\}$, and ${\cal L}\{\cdot \}$ denotes the Laplace transform. The parameters $v_1=1$, $v_2=4$, and $v_3=10$ are calculated from the data at the population level and kept fixed across the cohort. The static gain of transfer function \eqref{eq:lin_NMB} is normalized to one and $0<\alpha\le  0.1$ is the only patient-specific parameter. The PK model in \eqref{eq:lin_NMB} can be written in state space as
\begin{equation}                            \label{eq:1nl}
\dot{x}(t) =Ax(t)+Bu(t), \quad \bar y(t)=Cx(t),
\end{equation}
where the matrices are of the same structure as in \eqref{eq:matrices} and $a_1=v_1\alpha$, $a_2=v_2\alpha$, $a_3=v_3\alpha$, $g_1=v_1\alpha$, $g_2=v_2v_3\alpha^2$.

The PD part is modeled by a Hill function of order $\gamma$
\begin{equation}\label{eq:nonlin_NMB}
y=\varphi(\bar y)=\frac{100 C_{50}^\gamma}{ C_{50}^\gamma + {\bar y}^\gamma(t)}, \quad \gamma>0,
\end{equation}
where $C_{50}=3.2425$ $ \mu \mathrm{ g} \ \mathrm{ml}^{-1} $ is the drug concentration that yields 50\% of the maximum effect.
The output $y(t)$ $\lbrack \% \rbrack$ represents  the effect of the NMB agent and is measured by a train-of-four (ToF) neuromuscular monitor~\citep{MH06}. The maximal level of $y(t)=100\%$ corresponds to no NMB, i.e. before effective drug action,
when there is no drug in the patient's bloodstream. The point $x(0)=0$ thus provides the initial conditions for the controller.

In model~\eqref{eq:lin_NMB},~\eqref{eq:nonlin_NMB}, a patient's response to a drug dose is captured with a pair $(\alpha,\gamma)$.  The dataset used in this study includes individualized models of 48 patients estimated from clinical data obtained under closed-loop administration and is described in detail in~\cite{ML98}. Mathematical modeling  with model structure~\eqref{eq:lin_NMB},~\eqref{eq:nonlin_NMB} was first performed in~\cite{SWM12} and revisited in~\cite{RMM14} with a different estimation algorithm.

Standard PKPD models in pharmacometrics make use of more independent parameters than two. For instance, in~\cite{FLA01}, the PK part (the Sheiner model) is parametrized in four parameters and at least one independent parameter is required in the PD part. To individualize such a model, a high degree of exogenous (time/frequency) excitation is required  for accurate system identification. Besides, the input must drive the system across its full operating range so that the PD nonlinearity can be properly mapped and parameterized. The latter aspect is often missed in practice of PKPD identification and measurements are collected only for a limited number of fixed infusion rates.

\begin{figure}[ht]
\centering
\includegraphics[width=0.9\linewidth]{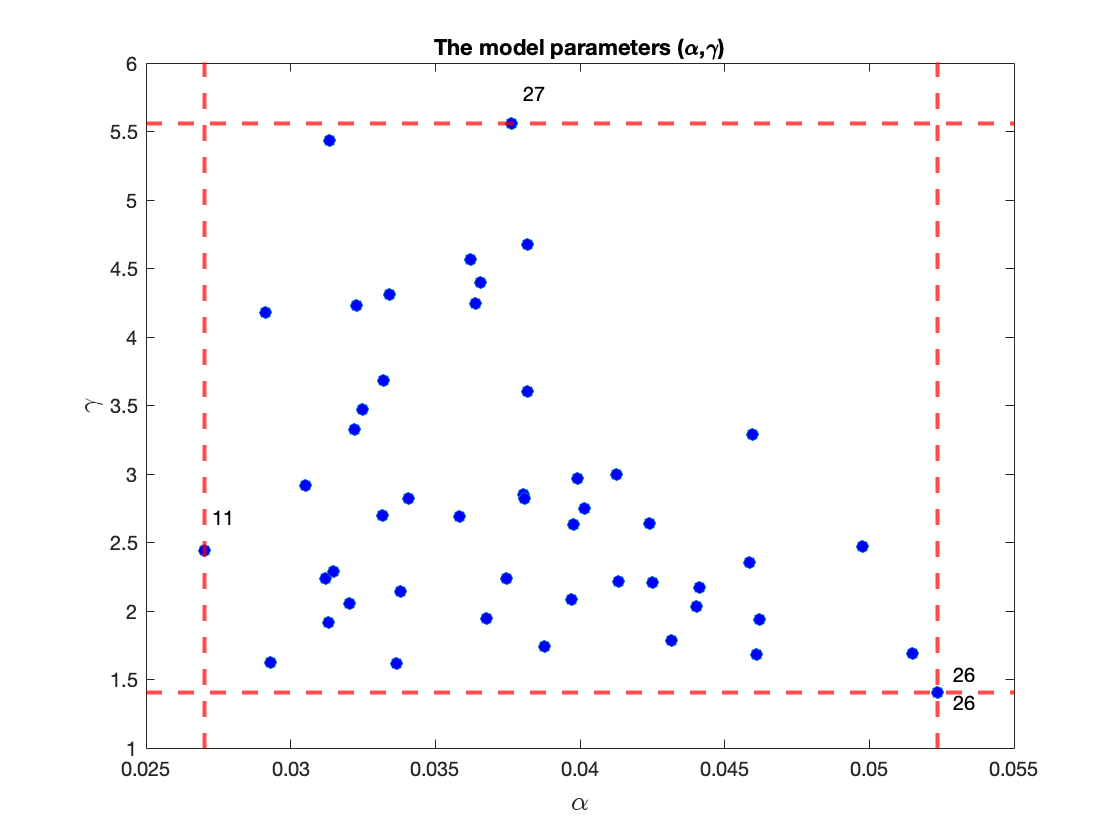}
\caption{The model parameter pairs in the dataset satisfy $\alpha_{\min}= 0.0270\le \alpha \le 0.0524=\alpha_{\max}$, $\gamma_{\min}=1.4030\le\gamma\le 5.5619=\gamma_{\max}$. The extreme parameter values are indicated by the Patient Identification Number (PIN). PIN~26 features two extreme values and corresponds to $(\alpha_{\max},\gamma_{\min})$.}\label{fig:alpha_gamma}
\end{figure}

The model parameter estimates for the patient  cohort are illustrated in Fig.~\ref{fig:alpha_gamma}. Apparently, high values of both $\alpha$ and $\gamma$ do not appear in the dataset and the data points are concentrated to the lower left part of the scatter plot. The population mean values are $\bar \alpha=0.0374$ and $\bar \gamma=2.6677$.

\subsection{Feasibility analysis}
Using {\it atracurium} in surgical practice, NMB is initiated with a bolus  dose of $400\text{--}500~\mathrm{\mu g/kg}$. Further into the procedure, maintenance doses of $80\text{--}200~\mathrm{\mu g/kg}$ are  administered every $15\text{--}25~\mathrm{min}$.
Dosing varies widely among patients and is individualized based on continuous monitoring of neuromuscular function, e.g. by a ToF monitor. From~\cite[Fig.~4]{SWM12}, the NMB depth is to be kept within the range $2\% \le y(t) \le 10\%$ throughout the surgery.  Then, this constitutes an output corridor control problem, cf.~\eqref{eq:corridor}, and can be solved by imposing a 1-cycle on the closed-loop system by means of impulsive feedback controller~\eqref{eq:1b}. Since the (zero) initial conditions of the closed-loop system lie far from the fixed point of the stationary solution, the fixed point has to be stable with sufficient attractivity~\cite{MPZ25}.

As pointed out before, accurate process modeling is essential in feedback dosing due to the patient safety reasons. It becomes even more important when pulse-modulated feedback is applied since the model is used to predict the PKPD dynamics not only over the time interval of one sample but for much longer time. Therefore, a method is required for evaluating whether a model estimated from data is feasible for calculating clinically acceptable doses.

A practical way of addressing the  feasibility of estimated models~\eqref{eq:lin_NMB},~\eqref{eq:nonlin_NMB} is to calculate what bolus doses and timing are required to maintain the model output within the prescribed corridor. This test helps to  exclude models whose use in impulsive feedback dosing would lead to dosing schemes that violate the recommended drug regimen based on established clinical practice. Indeed, with a patient-specific PKPD model expressed in terms of two constants (cf. \eqref{eq:lin_NMB}, \eqref{eq:nonlin_NMB}) and without any apparent connection to the purpose of medication, the model feasibility is difficult to evaluate. The results of Proposition~\ref{prop:corridor} and Theorem~\ref{th:solvability} support a straightforward procedure that allows to judge  whether or not a  PKPD model estimated from data is suitable for the calculation of sequential bolus doses. Notice that, in contrast to continuous infusion, impulsive controller \eqref{eq:1b} operates in the same terms that are used in manual bolus administration and, therefore, produces dosing schemes that are readily comparable to the manual mode.

The model feasibility evaluation procedure is as follows.
\begin{description}
\item[Step~1:] Set the output corridor limits in \eqref{eq:corridor} to $y_{\min}^*=2$, $y_{\max}^*=10$, according to the clinical recommendations.
\item[Step~2:] By inverting the PD function in \eqref{eq:nonlin_NMB}, calculate $\bar y_{\min}^*=\varphi^{-1}(y_{\max}^*)$, $\bar y_{\max}^*=\varphi^{-1}(y_{\min}^*)$.
\item[Step~3:] Given the matrices $A,B,C$ of the patient-specific model, solve the design equation
\begin{equation*}
\frac{z_{\max}(T)}{z_{\max}(T)\!-\!z_{\min}(T)} =\frac{\bar y_{\max}^*}{ \bar y_{\max}^*\! - \! \bar y_{\min}^*},
\end{equation*}
and obtain the value of $T^*$.
\item[Step~4:] Calculate the bolus dose
\begin{equation*}
    \lambda^*=\frac{\bar y_{\max}^*}{z_{\max}(T^*)}.
\end{equation*}
\item[Step~5:] Check whether the pair $(T^*,\lambda^*)$ (the dosing parameters) is within the clinically feasible interval.
\end{description}

In fact, the algebraic inverse of the PD function $\varphi(\cdot)$ is not necessary since the calculation of $\bar y$ from $y$ can be performed numerically.

\begin{figure}[ht]
\centering
\includegraphics[width=0.8\linewidth]{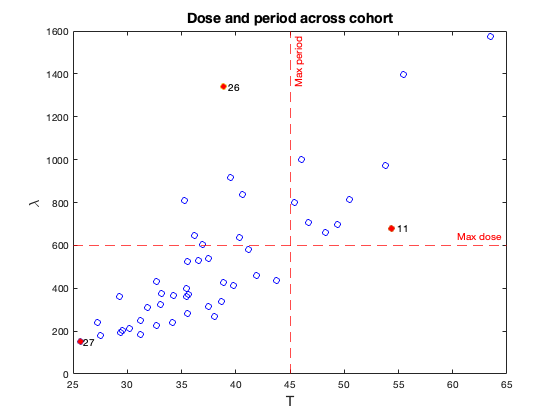}
\caption{Calculated values of $\lambda^*$ and $T^*$ across the cohort of patient models. The dashed lines show the highest clinically feasible values: $\lambda_{\max}=600~\mathrm{\mu g/kg}$, $T_{\max}=45~\mathrm{min}$. The cases exhibiting the extreme values of $\alpha$ and $\gamma$ (in red) are marked by PIN; cf. Fig.~\ref{fig:alpha_gamma}.}\label{fig:dose_T}
\end{figure}

    Calculations of $(T^*,\lambda^*)$ for all patients models in the cohort have been performed according to the procedure above and the results are summarized in Fig.~\ref{fig:dose_T}. It is instructive to examine the locations of the dosing parameters for the patient models with the extreme values of $\alpha$ and $\gamma$, i.e. PIN=11,26,27. Notably, the maximal value of $\gamma$ does not present a challenge (PIN=27) when it comes to output corridor control. In contrast, low values of $\alpha$ and $\gamma$ demand either excessively high drug doses or very long intervals between the dosing instants, to satisfy \eqref{eq:corridor}. Within the considered cohort, there is no model that yields a combination of low $\lambda^*$ and high $T^*$. This is also consistent with the least-squares nature of the estimator (the extended Kalman filter) employed in~\cite{SWM12} for model estimation.

\begin{figure}[ht]
\centering
\includegraphics[width=0.9\linewidth]{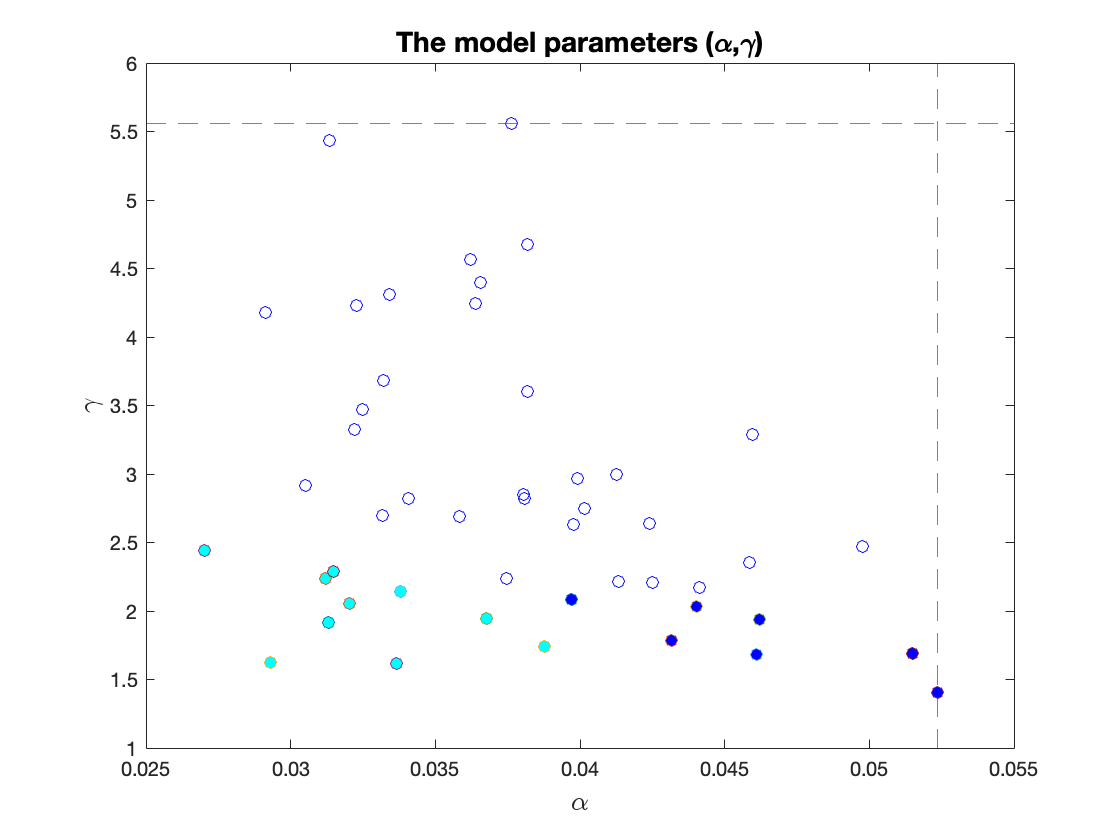}
\caption{The parameter pairs $(\alpha,\gamma)$ in the dataset. Infeasible models are marked with filled circles. Models with $\lambda_{\max}<\lambda^*$ and $T^*<T_{\max}$ are plotted in blue, and models with $\lambda_{\max}<\lambda^*$ and $T_{\max}<T^*$ are in cyan.}\label{fig:alpha_gamma_classified}
\end{figure}

To clarify the patterns in the  parameters $(\alpha, \gamma)$ resulting in infeasible patient models, the scatter plot in Fig.~\ref{fig:alpha_gamma} is repeated in Fig.~\ref{fig:alpha_gamma_classified}, but with the parameter pairs corresponding to infeasible models  highlighted by filled circles. A common property of all the infeasible models is low values of $\gamma$, i.e. the slope of the Hill function in \eqref{eq:nonlin_NMB} being too shallow. Apparently, a low value of $\gamma$ makes the model infeasible no matter what the value of $\alpha$ is. This can be interpreted as a low patient sensitivity to the drug resulting in prohibitively high doses of  atracurium required to achieve the desired effect, i.e. the output corridor defined by \eqref{eq:corridor}. The value of $\alpha$ is of less importance as it defines how fast the drug is eliminated from the organism.

There are several possible explanations or contributing reasons why some of the identified Wiener PKPD models do not agree with clinical dosing practices.
\begin{itemize}
    \item First of all, the mode of drug administration considered in the present paper, i.e., sequential boluses, is not the same as the one covered in the data underlying the modeling PKPD in~\cite{SWM12}, i.e. continuous infusion. By its sheer nature, impulsive control requires mathematical models that are accurate over a larger range of input amplitudes, whereas a controller stabilizing the closed-loop system in vicinity of a stationary point can be obtained based on a model that approximates the plant dynamics in this particular point.
    \item Further, as pointed out in~\cite{SWM12}, the plant excitation is typically low in drug infusion applications which inevitably leads to poor model parameter estimates. This is especially true when it comes  to estimating the model nonlinearity since the identification data reflect mostly the process operation under stationary conditions.
    \item The data underlying the used models were collected under feedback, see~\cite{ML98}. After the initial bolus, a proportional controller was applied  whereas a PID-controller was used in the maintenance phase of NMB. The closed-loop drug administration was not taken into account in the system identification, which typically leads to biased estimates, see~\cite{SS88}.
\end{itemize}

\section{Proofs of The Main Results}\label{sec:proof}

We start by establishing some properties of the functions $\tau_1=\tau_1(T)$ and $\tau_2=\tau_2(T)$ for $T>0$; the proof of Proposition~\ref{prop:corridor} will also be given in the next subsection.

\subsection{The Existence and Technical Properties of $\tau_1,\tau_2$}

The next lemma follows from the general theory of $T$-systems (Chebyshev systems)~\citep{KarlinStudden1966}.
\begin{lemma}\label{lem:chebysh} Let $\alpha_1,\ldots,\alpha_n\in\mathbb{R}$ be pairwise distinct and let $\sum_{i=1}^n|c_i|>0$, where $c_1,\ldots,c_n\in\mathbb{R}$.
Then the equation \[ \varrho(t)\triangleq\sum_{j=1}^nc_j\e^{t\alpha_j}=0 \] has no more than $n-1$ real solutions.
\end{lemma}

Lemma~\ref{lem:chebysh} entails the following property of the linear system~\eqref{eq:state-space},~\eqref{eq:matrices} impulse response.
\begin{prop}\label{prop:impulse}
The impulse response $g(t) = C \mathrm{e}^{tA} B$ of linear system~\eqref{eq:state-space},~\eqref{eq:matrices} attains its peak at some $t_{*}\in(0,\infty)$, increasing on $[0,t_{*})$ and decreasing on $(t_*,\infty)$.
\end{prop}
\begin{pf}
System~\eqref{eq:state-space} has a nonzero transfer function from $u$ to $y$; hence $g(t)$ and $\dot{g}(t)$ are nonzero linear combinations of the modal functions $\mathrm{e}^{-a_i t}$.
By Lemma~\ref{lem:chebysh}, $g$ has at most two stationary points where $\dot g=0$. One of these stationary points is $t=0$, because
$\dot{g}(0) = C A B = 0$ in view of~\eqref{eq:matrices}. Since $g(0)=CB=0$ and $g(t) \to 0$ as $t \to \infty$, the function $g$ cannot be strictly increasing on $(0,\infty)$; in other words, $\dot g$ cannot be sign-preserving. Hence, another stationary point $t_* \in (0, \infty)$ exists.
By noticing that $\ddot g(0)=CA^2B>0$, it follows that $\dot g(t)>0$ (i.e., $g(t)$ increases) for $0<t<t_*$ and $\dot g(t)<0$ (i.e., $g(t)$ decreases) for $t>t_*$. $\blacksquare$
\end{pf}

Proposition~\ref{prop:corridor} can now be proved similarly.
\paragraph*{\textbf{Proof} of Proposition~\ref{prop:corridor}}
\vskip-5mm
For $T>0$ being fixed, denote $X_T:=(\e^{-TA}-I)^{-1}B$ and $z(\tau):=z(\tau,T)$.
Then, $z(T)=CX_T$. Furthermore, $z(0)=z(T)$, since $CB=0$ and
\[
z(0)-z(T)=C(I-\e^{TA})(I-\e^{TA})^{-1}B=CB=0.
\]
Similarly, using the relation $CAB=0$, the derivative
\[
\dot{z}(\tau)= C\e^{\tau A}A(I-\e^{AT})^{-1}B
\]
satisfies the periodicity condition $\dot{z}(0)=\dot{z}(T)=CAX_T<0$. The latter inequality holds, because $AX_T$ is a negative vector~\cite[Proposition~3]{MPZh23}.
Therefore, $z(0)=z(T)$ is neither the minimum nor the maximum value of $z(\cdot)$ on the interval $[0, T]$. Since, by the Weierstrass theorem, $z(\cdot)$ attains its minimum and maximum values on $[0, T]$, there exist at least two extremal points $\tau_1,\tau_2$ with $0<\tau_1<\tau_2<T$ such that $\dot{z}(\tau_i) = 0$, or,
equivalently, $\tau = \tau_i$ is a solution of~\eqref{eq:roots_dy}. According to Lemma~\ref{lem:chebysh}, equation~\eqref{eq:roots_dy} cannot have more than two real roots since its left-hand side (for fixed $T$) is a nonzero linear combination of $\e^{-a_i\tau}$, $i=1,2,3$. Hence, $\tau_1$ and $\tau_2$ are the only points where $\dot{z}$ can change sign.
Recalling that $\dot z(0)=\dot z(T)<0$, it follows that $\dot{z}$ is negative on $(0, \tau_1)$ and $(\tau_2, T)$, and positive on $(\tau_1, \tau_2)$.
Consequently, $\tau_1$ is the point of minimum, and $\tau_2$ is the point of maximum, which completes the proof of~\eqref{eq:y_max} and~\eqref{eq:y_min} since $y(\tau)=\lambda z(\tau,T)$ for all $\tau\in[0,T]$ and the fixed point is $X=\lambda X_T$. $\blacksquare$

Notice that, in accordance with Proposition~\ref{prop:impulse}, one has $\dot g(t)=CA\e^{tA}B<0$ for $t>t_*$, in particular,
\[
\frac{\partial z(\tau,T)}{\partial\tau}=CA\e^{\tau A}(I-\e^{TA})^{-1}B=\sum_{k=0}^{\infty}CA\e^{(\tau+kT)A}B<0
\]
for any $\tau\geq t_*$. This entails the following corollary.
\begin{cor}
For any $T>0$, one has $0<\tau_1(T)<\tau_2(T)<t_*$. Furthermore, the following limit relations hold:
\begin{equation}\label{eq:limits}
\begin{gathered}
\tau_1(T)\xrightarrow[T\to\infty]{} 0,\quad \tau_2(T)\xrightarrow[T\to\infty]{} t_*,\\
z_{\min}(T)\xrightarrow[T\to\infty]{} 0,\quad z_{\max}(T)\xrightarrow[T\to\infty]{} g(t_*).
\end{gathered}
\end{equation}
\end{cor}
\begin{pf}
The first statement follows immediately from Propositions~\ref{prop:impulse} and~\ref{prop:corridor}: recall that $\dot g(t)\geq 0$ for $t\in[\tau_1,\tau_2]$, which means that
$[\tau_1,\tau_2]\cap [t_*,\infty)=\emptyset$, i.e., $\tau_2(T)<t_*$ for all $T$.
Hence, for $T>t_*$, the minimum and maximum of $z(\tau,T)$ on the intervals $[0,T]$ and $[0,t_*]$ are identical.
The relations~\eqref{eq:limits} now follow from the fact that $z(\tau,T)$ converges to $g(\tau)$ uniformly in $\tau\in[0,t_*]$ as $T\to\infty$, i.e.,
\[
\max_{0\leq\tau\leq t_*}|z(\tau,T)-g(\tau)|\xrightarrow[T\to\infty]{}0,
\]
and from the fact that $g(t)$ attains a unique global minimum (equal to $0$, at $t=0$) and a unique global maximum (at $t=t_*$) on $[0,t_*]$.
\end{pf}

It can be easily shown that $z_{\min}$ and $z_{\max}$ tend to $\infty$ as $T\to 0$, however, $z_{\max}-z_{\min}$ remains bounded  (in fact, simulations suggest that this function vanishes as $T\to 0$; see Fig.~\ref{fig:num_den_design}). We formulate the following corollary.
\begin{cor}
The following limit relations hold:
\begin{equation}\label{eq:limits1}
\begin{gathered}
z_{\min}(T)\xrightarrow[T\to 0]{} +\infty,\quad z_{\max}(T)\xrightarrow[T\to 0]{} +\infty,\\
\limsup_{T\to 0}\;(z_{\max}(T)-z_{\min}(T))<\infty.
\end{gathered}
\end{equation}
\end{cor}
\begin{pf}
Notice that $(I-\e^{TA})^{-1}=-\dfrac{1}{T}A^{-1}+O(1)$ as $T\to 0$. Since $(-CA^{-1}B)=\dfrac{g_2g_1}{a_1a_2a_3}>0$, one has
\[
z_{\max}(T)>z(0,T)=-\dfrac{1}{T}(CA^{-1}B)+O(1)\xrightarrow[T\to 0]{}\infty.
\]
The mean value theorem guarantees that
\[
0<z(\tau_2,T)-z(\tau_1,T)=(\tau_2-\tau_1)\left.\frac{\partial}{\partial\tau}z(\tau,T)\right|_{\tau=\xi},
\]
where $0<\tau_1<\xi<\tau_2<T$ (the point $\xi=\xi(T)$ depends on $T$). Since $0<\tau_2-\tau_1<T$ and, in view of~\eqref{eq:roots_dy}, the derivative is $O(1/T)$ as $T\to 0$, the right-hand side remains bounded as $T\to 0$. $\blacksquare$
\end{pf}

Introduce the function
\[
\hat z(\tau,T)= \begin{cases} z(\tau,T), & \tau\leq T,\\ z(T,T), & \tau>T. \end{cases}
\]
It follows from Proposition~\ref{prop:corridor} that $\tau_1(T)$ and $\tau_2(T)$ are the unique argmin and argmax, respectively, of $\hat z(\cdot,T)$ on $[0,\infty)$. Since $\hat z$ is continuous on $[0,\infty)\times(0,\infty)$, Berge's Maximum Theorem~\citep{Ok2007} yields the following.
\begin{cor}\label{cor:contin}
The functions $\tau_1$, $\tau_2$, $z_{\min}$, $z_{\max}$ are continuous on $(0,\infty)$.
\end{cor}

\subsection{Proof of Theorem~\ref{th:solvability}}

Since $z_{\min}$ and $z_{\max}$ are continuous (Corollary~\ref{cor:contin}), the function $\Psi(\cdot)$ in~\eqref{eq:design_T} is continuous. According to~\eqref{eq:limits} and~\eqref{eq:limits1}, its limits as $T\to 0$ and $T\to\infty$ are, respectively, $\infty$ and $1$. Hence, this function takes all values in the interval $(1,\infty)$, in particular, the equation~\eqref{eq:design_T} has a solution.
The remaining statements follow from Proposition~\ref{prop:corridor}.
$\blacksquare$

\section*{Conclusions}
The solvability of the output corridor impulsive control problem is investigated. It is proven that for the linear positive plant  comprising a cascade of three first-order blocks, a stationary 1-cycle satisfying any prescribed positive output corridor always exists under  pulse-modulated feedback. The obtained result is shown to be useful in feasibility analysis of identified minimal  patient-specific pharmacokinetic-pharmacodynamic models. The algorithm for calculating the exact upper and lower bounds of the stationary periodic solution with one firing of the impulsive feedback on the least period (1-cycle) allows solving the converse problem. Then, for a desired output corridor and given plant model, the drug dose and inter-dose period can be calculated and compared to clinically practiced regimen.
\bibliography{refs}
\end{document}